\documentclass[twocolumn,showpacs,preprintnumbers,amsmath,amssymb,prb]{revtex4}
\usepackage{epsfig}
\usepackage{graphicx}
\usepackage{dcolumn}
\usepackage{bm}
\usepackage{color} 

\begin{document}

\title{Electronic inhomogeneity in EuO: Possibility of magnetic polaron states}

\author{Shin-ichi Kimura}
 \altaffiliation[Electronic address: ]{kimura@ims.ac.jp}
\affiliation{UVSOR Facility, Institute for Molecular Science, Okazaki 444-8585, Japan}
\affiliation{School of Physical Sciences, The Graduate University for Advanced Studies (SOKENDAI), Okazaki 444-8585, Japan}
\author{Hidetoshi Miyazaki}
\affiliation{Graduate School of Engineering, Nagoya University, Nagoya 464-8603, Japan}
\affiliation{UVSOR Facility, Institute for Molecular Science, Okazaki 444-8585, Japan}
\author{Takafumi Mizuno}
\author{Takuya Iizuka}
\affiliation{School of Physical Sciences, The Graduate University for Advanced Studies (SOKENDAI), Okazaki 444-8585, Japan}
\author{Toshiharu Takahashi}
\affiliation{Research Reactor Institute, Kyoto University, Osaka 590-049, Japan}
\author{Takahiro Ito}
\affiliation{UVSOR Facility, Institute for Molecular Science, Okazaki 444-8585, Japan}
\affiliation{School of Physical Sciences, The Graduate University for Advanced Studies (SOKENDAI), Okazaki 444-8585, Japan}
\date{\today}
\begin{abstract} 
We have observed the spatial inhomogeneity of the electronic structure of a single-crystalline electron-doped EuO thin film with ferromagnetic ordering by employing infrared magneto-optical imaging with synchrotron radiation.
The uniform paramagnetic electronic structure changes to a uniform ferromagnetic structure via an inhomogeneous state with decreasing temperature and increasing magnetic field slightly above the ordering temperature.
One possibility of the origin of the inhomogeneity is the appearance of magnetic polaron states.
\end{abstract}

\maketitle
%
%
Europium monoxide (EuO) is a ferromagnetic semiconductor with a Curie temperature ($T_{\rm C}$) of around 70~K.~\cite{Mauger1986}
With excess Eu electron doping or the substitution of Gd$^{3+}$ or La$^{3+}$ for Eu$^{2+}$-ions, $T_{\rm C}$ increases up to 150~K and the electrical resistivity drops twelve orders of magnitude below $T_{\rm C}$.~\cite{Oliver1972}
Since the magnetic moment originates from the local Eu$^{2+}$~$4f^7$ electrons, the electron-doped EuO has larger magnetic moments than colossal magneto-resistance manganites, which exhibit a similar insulator-to-metal transition at $T_{\rm C}$.~\cite{Asamitsu1995}
Therefore, the electron-doped EuO is attracting attention as a next generation functional material for a spintronics device.~\cite{Schmehl2007}

There are two theories to explain the origin of the increasing $T_{\rm C}$ and the insulator-to-metal transition in electron-doped EuO.
One is a magnetic polaron scenario in which heavy carriers resulting from the exchange interaction between the local $4f$ electrons and carriers, namely $cf$ interaction, are trapped by donor states and polarize the $4f$ electrons around the carriers.~\cite{Kasuya1968,Nagaev1983,Umehara1987,Umehara1989,Yu2005,Yu2006}
The magnetic polaron states make a percolative state between ferromagnetic and paramagnetic domains.
Such phase separation has been predicted theoretically, but has never been directly observed.
Another scenario is where the conduction band is split by the ferromagnetic transition and hybridizes with the donor state.~\cite{Oliver1972,Emin1986,Emin1987,Hillery1988,Arnold2007}
Both theories can explain the behavior of the electrical resistivity and magnetic susceptibility.
The difference is whether or not the phase separation appears.
With the magnetic polaron scenario, there is phase separation around $T_{\rm C}$ in which the ferromagnetism and paramagnetism coexist.~\cite{Yu2005,Yu2006}
If phase separation is observed, the magnetic polaron scenario accounts for the origin of the phase transition from paramagnetic insulator to ferromagnetic metal.

At the ferromagnetic transition in EuO, the Eu~$5d$ conduction band shifts to the Fermi level owing to the strong $cf$ interaction.
In addition, it is known that the exciton absorption edge of the Eu~$4f\rightarrow5d$ transition at around 10000~cm$^{-1}$ shifts to the lower energy side.~\cite{Mitani1975,Sakai1977}
The energy shift is a good probe for the ferromagnetic transition.
In this paper, to confirm the existence of the magnetic polaron state, we investigate the temperature and magnetic field dependence of the real space distribution of the ferromagnetic electronic structure using the absorption edge shift of a slightly Eu-rich EuO thin film.
As a result, we found that spatial inhomogeneity with sub-$\mu$m size ferromagnetic domains in both the temperature and magnetic field dependences near $T_{\rm C}$.
The origin of the creation of the domains could be magnetic polaron states.

%
%
Single-crystalline EuO thin films have been grown by using molecular beam epitaxy (MBE).
To obtain high-quality single-crystalline films, we evaporated Eu onto BaO buffered SrTiO$_3$ substrates at 350~$^{\circ}$C under an oxygen pressure of 8~$\times$~10$^{-6}$~Pa.~\cite{Iwata2000-1,Iwata2000-2}
The epitaxial growth of single-crystalline EuO thin films with 1~$\times$~1 EuO $(100)$ patterns has been checked with LEED and RHEED methods.~\cite{Miyazaki2008}
The sample thickness was 100~nm as detected with a quartz resonator sensor.

\begin{figure}[t]
\begin{center}
\includegraphics[width=0.4\textwidth]{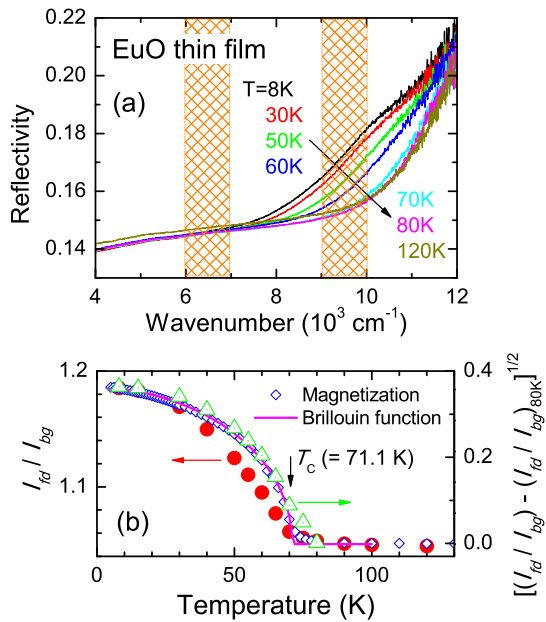}
\end{center}
\caption{
(Color online) (a) Temperature dependence of the reflectivity spectrum $R(\omega)$ of an EuO thin film in the 4000~--~12000~cm$^{-1}$ wave number range.
(b) The temperature dependence of the intensity ratio between the absorption edge of the exciton of the Eu~$4f\rightarrow5d$ transition ($I_{fd}$) integrated at 9000~--~10000~cm$^{-1}$ and the background intensity ($I_{bg}$) at 6000~--~7000~cm$^{-1}$ (solid circle, left axis) and the function of $[(I_{fd}/I_{bg})-(I_{fd}/I_{bg})_{80 K}]^{1/2}$ (open triangle, right axis).
The temperature dependence of the magnetization (open diamond) and the fitting curve (solid line) using the Brillouin function are also plotted.
See text for details.
}
\label{Refl}
\end{figure}

The infrared reflectivity spectra [$R(\omega)$] and spatial images with unpolarized light were obtained at the infrared magneto-optical station of the beam line 43IR of a synchrotron radiation facility, SPring-8, Japan.~\cite{Kimura2001,Kimura2003}
The infrared $R(\omega)$ imaging was performed in the 6000 to 12000~cm$^{-1}$ wave number range with 10~cm$^{-1}$ resolution at different temperatures from 40 to 80~K in magnetic fields up to 5~T.
The measurements were performed only with increasing temperature and increasing magnetic field strength.
To acquire the spatial imaging data, a total of 1681 spectra were obtained in a 200~$\times$~200~$\mu$m$^{2}$ region in steps of 5~$\mu$m with a spatial resolution of better than 5~$\mu$m.
The accumulation time of one image takes about 3 hours.
The spatial images and the temperature dependence of the reflectivity spectrum were plotted using the intensity ratio of the absorption edge of the exciton of the Eu~$4f \rightarrow 5d$ transition ($I_{fd}$) integrated over 9000~--~10000~cm$^{-1}$ to the background intensity ($I_{bg}$) in the 6000~--~7000~cm$^{-1}$ range as shown in Fig.~\ref{Refl}(a).
To check the reproducibility of the spatial images, the same measurement was repeated.
A good reproducibility was observed, indicating that the inhomogenity is static during measurements.

%
%
The temperature dependence of $R(\omega)$ of a single-crystalline EuO thin film in an area of about 4~mm$^2$ is shown in Fig.~\ref{Refl}(a).
The figure indicates that the absorption edge clearly shifts to the lower energy side with increasing temperature.
The temperature dependence of $I_{fd}/I_{bg}$ is plotted in Fig.~\ref{Refl}(b).
The temperature dependence of the magnetization measured with a SQUID magnetometer (Quantum Design MPMS-7) and the fitting curve using the Brillouin function with $J=7/2$ are also plotted in the figure.~\cite{AM1976}
$I_{fd}/I_{bg}$ rapidly increases below $T_{\rm C}$~=~71.1~K as the magnetization increases.
The $T_{\rm C}$ is slightly higher than the previous result ($\sim$~69~K).~\cite{Mauger1986}
The higher $T_{\rm C}$ indicates a slight excess of Eu.
Since the reflectivity intensity is proportional to the square of the modulus of the electric field, and since the magnetization should be proportional to the electric field of the reflected light (minus the background light), the square root of the properly normalized intensity is expected to follow the behavior of the magnetization.~\cite{Zvezdan1997}
The function of $[(I_{fd}/I_{bg})-(I_{fd}/I_{bg})_{80~K}]^{1/2}$ is also plotted in Fig.~\ref{Refl}(b) and is in good agreement with the magnetization curve.
Therefore the energy shift of the absorption edge strongly relates to the ferromagnetic ordering.

\begin{figure}[t]
\begin{center}
\includegraphics[width=0.4\textwidth]{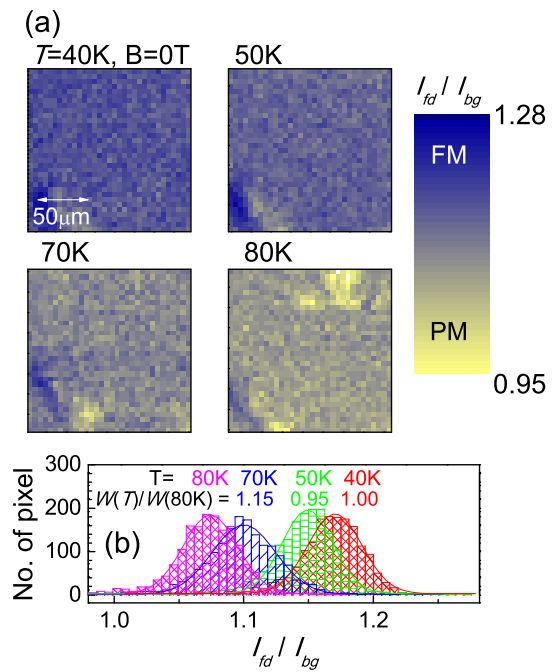}
\end{center}
\caption{
(Color online) (a) Temperature dependence of the spatial image of the intensity ratio ($I_{fd}/I_{bg}$) of an EuO thin film in a magnetic field of $B$~=~0~T.
(b) Statistical distributions of $I_{fd}/I_{bg}$ derived from (a).
Gaussian fittings to the histograms are also plotted using solid lines and the widths divided by the width at 80~K [$W(T)/W({\rm 80~K})$] are also shown.
}
\label{Tdep}
\end{figure}
To obtain the temperature dependence of the spatial distribution of the ferromagnetic domains, we plotted the spatial image of $I_{fd}/I_{bg}$ as shown in Fig.~\ref{Tdep}(a).
The figure shows that the whole sample is monotonically ferromagnetic at $T$~=~40 and 50~K.
With increasing temperature, $I_{fd}/I_{bg}$ in the whole sample decreases and then a portion changes to the paramagnetic state because $I_{fd}/I_{bg}$ becomes small.
The phase separation mainly originates from the inhomogeneity of the sample itself, which is the same as that observed in organic conductors.~\cite{Nishi2005,Nishi2007}
Then, to investigate the additional change in the sample inhomogeneity, the statistical distribution of $I_{fd}/I_{bg}$ in the whole sample is plotted in Fig.~\ref{Tdep}(b).
The distribution width at each temperature divided by that at 80~K [$W(T)/W({\rm 80~K})$] is also shown in the figure.
From the figure, the distribution width of $I_{fd}/I_{bg}$ does not change greatly but the peak wave number changes with temperature.
However, the width at 70~K increases by 15~\%.
This is evidence of another effect in addition to the sample inhomogeneity.
Generally, at the magnetic ordering temperature, the local electronic structure in materials simultaneously changes due to the long-range magnetic ordering.~\cite{Kimura2002}
However, when magnetic polaron states appear near $T_{\rm C}$, the phase separation between the ferromagnetic and paramagnetic domains must appear.
The phase separation makes the increase of the distribution width.
Therefore, the increase of the distribution width at 70~K can be attributed to the appearance of the magnetic polaron state.

\begin{figure}[t]
\begin{center}
\includegraphics[width=0.4\textwidth]{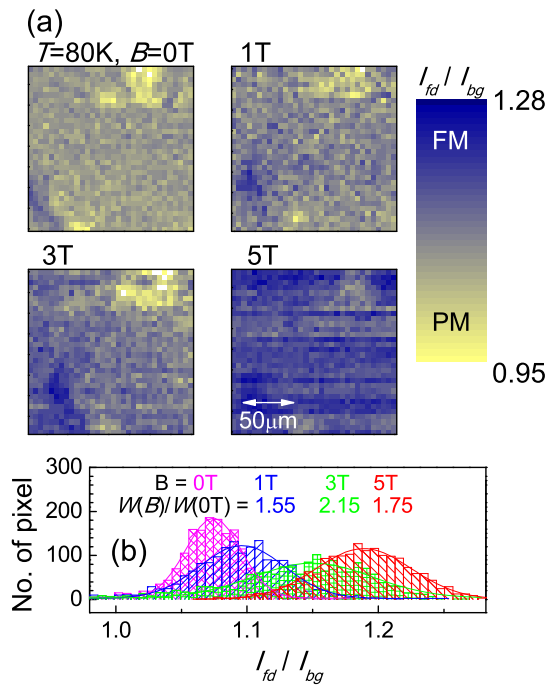}
\end{center}
\caption{
(Color online) (a) Magnetic field dependence of $I_{fd}/I_{bg}$ of an EuO thin film at $T$~=~80~K.
(b) Statistical distributions of $I_{fd}/I_{bg}$ derived from (a).
Gaussian fittings to the histograms are also plotted using lines and the widths divided by the width at 0~T [$W(B)/W({\rm 0~T})$] are also shown.
}
\label{Bdep}
\end{figure}
Such additional inhomogeneity also appears in the magnetic field dependence.
To investigate the phase separation in the magnetic field induced paramagnetic-to-ferromagnetic transition, we performed infrared imaging under magnetic fields at 80~K as shown in Fig.~\ref{Bdep}(a).
The figure indicates that the paramagnetic state at 0~T becomes ferromagnetic as the magnetic field increases.
Similar magnetic-field-induced insulator-to-metal transition has been observed in manganites in spite of the different origin.~\cite{Fath1999}
The spatial distribution, which we analyzed in the same way as in Fig.~\ref{Tdep}(b), is plotted in Fig.~\ref{Bdep}(b).
The figure shows that the distribution width normalized by that at 0~T [$W(B)/W({\rm 0~T})$] exhibits an approximately twofold increase at 3~T and then decreases at 5~T.
Such anomalous temperature-dependent width can be explained in terms of the magnetic polaron scenario, in which a large magnetic polaron is created by applying a magnetic field.
In other words, the phase transition from the uniform paramagnetic state to the uniform ferromagnetic state via the inhomogeneous magnetic polaron state appears with increasing magnetic field at a temperature slightly higher than $T_{\rm C}$.
This result is consistent with the NMR result, where magnetic inhomogeneities are formed when the temperature is increased.~\cite{Comment2005}

We use these results as a basis for discussing the domain size of the magnetic polaron state in EuO.
The spatial resolution of about 5~$\mu$m used in this work is larger than this domain size.
However, changes in the spatial distribution width can be detected.
In such cases, the domain size is about one-tenth of the spatial resolution of the used microscope, {\it i.e.}, about several 100~nm.
In a theoretical study, the domain size of the magnetic polaron is reported to be several nm.~\cite{Umehara1989}
However, this is an ideal case where the magnetic polaron size is uniform.
In the present case, the excess Eu ions are regarded to exist at random because the $I_{fd}/I_{bg}$ image in Figs.~\ref{Tdep}(a) and \ref{Bdep}(a) is not monochromatic even at 40~K.
Therefore a larger domain size of the magnetic polaron state is estimated to be several 100~nm.

Recently, the percolated insulator-metal phase separation with domain sizes of several 100~nm to several $\mu$m of other strongly correlated electron systems; one example is VO$_2$ at the Mott transition boundary, which has been observed by using a scattering-type scanning near-field infrared microscope,~\cite{Qazil2007} and another one is manganites due to the charge-ordering, which has been observed by using a spatially resolved photoemission~\cite{Sarma2004} and by a scanning tunneling spectroscopy.~\cite{Fath1999}
The expected domain size of EuO is coincidentally similar to those of VO$_2$ and the manganites, even though the origin of the phase separation is different.
To probe the electronic inhomogeneity in the sub-$\mu$m to $\mu$m domain size gives us new information of physical properties of materials.

%
%
To summarize, we measured the temperature and magnetic field dependences of the spatial distribution of the absorption edge of electron-doped EuO.
In terms of temperature dependence, we observed inhomogeneity at 70~K, which is close to $T_{\rm C}$.
As regards the magnetic field dependence at 80~K, namely slightly above $T_{\rm C}$, the spatial distribution width increases up to 3~T and decreases at 5~T, which indicates that the spatial inhomogeneity depends on the magnetic field.
One possible explanation of the temperature and magnetic field dependence of the spatial distribution width is the creation of the magnetic polaron state during the phase transition of the paramagnetic state to the ferromagnetic state.

%
%
We would like to thank staff members at BL43IR of SPring-8 for their technical supports.
This work was performed at SPring-8 with the approval of the Japan Synchrotron Radiation Research Institute (Proposal Nos.~2007A1253, 2007B1074) and was partly supported by a Grant-in-Aid of Scientific Research (B) (No.~18340110) from MEXT of Japan.


\end{document}